# Global Regularity and Individual Variability in Dynamic Behaviors of Human Communication

Jonathan J. H. Zhu
Dept. of Media and Communication
City University of Hong Kong
Hong Kong, China
j.zhu@cityu.edu.hk

Tai-Quan Peng
Web Mining Lab, Dept. of Media and Communication
City University of Hong Kong
Hong Kong, China
taiqpeng@cityu.edu.hk

*Abstract*—**A new model, called "Human Dynamics", has been recently proposed that individuals execute activities based on a perceived priority of tasks, which can be characterized by a power-law distribution of waiting time between consecutive tasks (Barabási, 2005). This power-law distribution has been found to exist in diverse human behaviors, such as mail correspondence, e-mail communication, webpage browsing, video-on-demand, and mobile phone calling. However, the pattern has been observed at the global (i.e., aggregated) level without considering individual differences. To guard against ecological fallacy, it is necessary to test the model at the individual level. The current study aims to address the following questions: Is the power-law uniform across individuals? What distribution do individual behaviors follow? We examine these questions with a client log file of nearly 4,000 Internet users' web browsing behavior and a server log file of $2.3 \times 10_6$ users' file sharing behaviors in a P2P system. The results confirm the human dynamic model at the aggregate-level both in webpage browsing and P2P usage bahevoirs. We have also found that there is detectable variability across the individuals in the decaying rate (i.e., the exponent γ) of the power-law distribution, which follows well-known distributions (i.e., Gaussian, Weibull, and log-normal).**

## I. INTRODUCTION

It was assumed in previous studies that human actions were performed at a constant rate, which was approximated with a Poisson distribution of inter-event time between two consecutive behaviors. The assumption was made due to the lack of long-term standard database about human activities [1, 2]. Rapid developments of information and communication technology, which accumulates large datasets recording selected human behaviors, makes it possible to question the validity of Poisson assumption.

In [3], Barabási found that the distribution of inter-event time in email communications had a bursty nature, which can be fitted with a power-law form $p(\tau) \sim \tau^{\gamma}$, where the exponent was equal to -1. The model of human dynamics developed in [3] has several advantages. First, it has adequate predictive power. The power-law property has been observed in diverse human behavior domains, such as surface mail communication [4], webpage browsing [5-8], social network sites [9], online games [10], video-on-demand [11], and mobile phone usage [2, 12, 13]. Second, it is parsimonious. The major parameter in the model was the exponent γ. Third, it has high generaliability because the major parameter in the model is scale-free. Finally, it is heuristically provocative. An ad hoc argument was proposed in [3] to explain the origin of bursty nature of human behaviors, which argued that the bursty nature of human dynamics is a consequence of queuing process driven by human decision making, while the previously-held assumption of Poisson distribution is a consequence of random process. Two universality class of human dynamics were further proposed in [1] based on empirical studies of human dynamics model proposed in [3], which classified human dynamics to two queuing models: one with γ = -3/2 assumes there are no limitations on the number of tasks an individual can handle at any time, while the other with γ = -1 assumes that there are limitations on the queuing length.

However, is the universality class really universal to all individuals? Most of those empirical studies about human dynamics were conducted at the aggregate level (i.e., by pooling all users together) assuming that individuals in the society have the average characteristics of the whole population at large. This will lead to the problem of ecological fallacy [14] which means erroneously drawing conclusions about individuals based solely on the observation of groups.

Specifically speaking, it is theoretically significant to systematically examine if the power-law regularity observed at the aggregate level can be generalized to the individual level. Our research questions in the current study are: (1) is the power-law distribution observed at the aggregate level uniform (i.e., with the same decision making mechanism) across all individuals? (2) If not, how much the individuals differ and what the differences look like?

## II. METHOD

The data used in the study come from two sources. The first dataset is a client log file which records 3,703 individuals' webpage browsing behaviors at home in four weeks spread out in 2002-2003 or 2003-2004. Those 3,703 individuals were randomly sampled in Hong Kong from 2002 to 2004. In total, $4.9 \times 10_6$ records are included in the dataset, with about 1,300 visits per user over 4 weeks or 46 visits per user per day on average. Each record consists of four elements: a user ID, a requested URL, the starting time of webpage request, and the end time of the webpage request.

The second dataset is a server log file of users downloading documents from each other through a P2P system (Maze developed by Peking University in China) from November 2008 to February 2009. There are $2.3 \times 10^5$ users in the dataset with about $1.0 \times 10^7$ records in the dataset. Each record consists of four elements: a source user ID (i.e., who downloads documents from others), a target user ID (i.e., whose documents are downloaded by others), the date of the downloading behavior, number of documents transferred between source user and target user on the day. In total, $3.7 \times 10^7$ documents are transferred in the four months and 165 documents per user on average.

Six operational definitions of human dynamics are developed in the current study, with four of them for webpage browsing behavior and remaining two for P2P usage. To measure human dynamics in webpage browsing, four operational definitions are developed: (a) inter-event time which is the time interval between two consecutive webpage requests; (b) preference of websites which is the time duration an user spend on a website per day; (c) stickiness to website which is the number of requests an user click on a website per day; (d) diversity of behaviors which is the number of websites an user visits per day. Two operational definitions are developed to measure human dynamics in P2P usage: (e) indegree which is the number of documents downloaded by others per day; (f) outdegree which is the number of documents downloaded from others per day.

### III. FINDINGS

#### A. Global Regularities of Human Dynamics

Human dynamics in webpage browsing and P2P usage is confirmed at the aggregate level. As shown in Figure 1, the distributions of inter-event time, preference of websites, stickiness to websites, and diversity of behaviors, based on the aggregated data of all users, can be well fitted by a power-law form $p(\tau) \sim \tau^{\gamma}$, where the exponents γ are -1.881, -1.413, -1.878, and -2.237 for four measurements of human dynamics in webpage browsing.

The heavy tailed nature of the observed inter-event time distribution has clear visual signature. Indeed, it implies that Internet users' webpage browsing has a bursty character: short time intervals with intensive activity bursts are separated by long periods of no activity. The heavy tailed nature of other three distributions in webpage browsing has also their own implications. The heavy-tailed distributions of preference of websites suggest that Internet users will spend very limited time on most of the websites while spend a lot of time on some specific websites. The heavy-tailed distribution of stickiness to websites implies that Internet users will click most of the websites one or two times per day. The heavy-tailed distribution of diversity of behaviors implies that Internet users will visit a very limited number of websites per day.

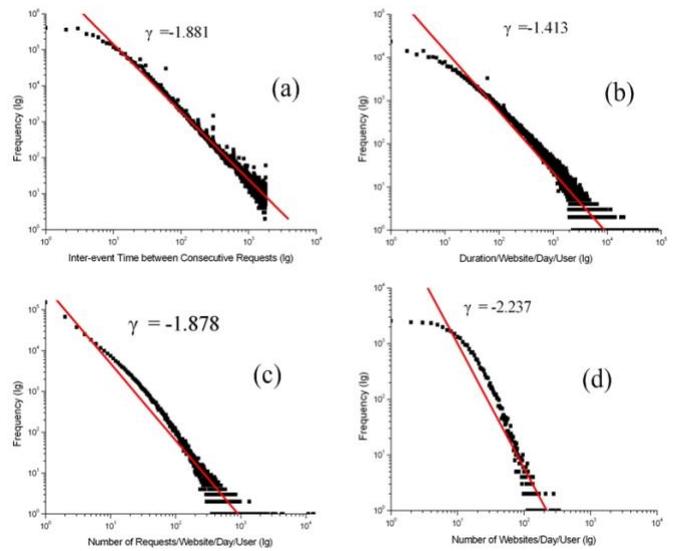

Figure1. Log-log plots of (a) distribution of inter-event time, (b) distribution of preference of websites, (c) distribution of stickiness to websites, (d) distribution of diversity of behaviors. The black dots represent the empirical data, and the red lines are the linear fittings.

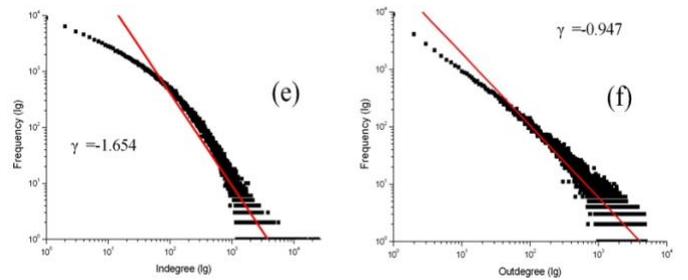

Figure 2 Log-log plots of (e) distribution of indegree and (f) distribution of outdegree. The black dots represent the empirical data, and the red lines are the linear fittings.

As shown in Figure 2, the distributions of indegree and outdegree in P2P usage, based on the aggregated data of all users, can also be well fitted by a power-law form $p(\tau) \sim \tau^{\gamma}$, where the exponents γ are -1.654 and -0.947, respectively. This implies that individuals' file sharing behavior in P2P usage also has a bursty character. They will conduct downloading/uploading behaviors with intensive bursts which are separated by long periods of no activity.

The exponents γ detected in P2P usage at the aggregate level are consistent with the two universality class of human dynamics in [1], respectively. The distribution of indegree referring to the documents downloaded by others from a user is consistent with the queuing model assuming no limitations on the number of tasks, while the distribution of outdegree referring to the documents downloading from others by a user is consistent with the queuing model assuming limitations on the number of tasks. This is in accordance with our intuitive expectation. Users of P2P system will seldom set limitations on number of documents downloaded by others. Therefore, the distribution of indegree will have an exponent close to -1.5 (in the study γ = -1.654). However, due to physical (e.g., limited Internet connection speed, inadequate CPU) and psychological constraints (e.g., limited cognitive capacity, limited needs), users of P2P system will intentionally or unintentionally set limitations on the number of documents downloaded from others. Therefore, the exponent of distribution of outdegree will approximate -1 (in the study γ = -0.947).

*B. Individual Variability on γ*

To investigate if the power-law distribution detected at the aggregate level can be confirmed at the individual level, we fit each individual with a power-law form $p(\tau) \sim \tau^\gamma$ in two datasets. Specifically speaking, each individual gets four γs in webpage browsing and two γs in P2P usage because four measurements of human dynamics in webpage browsing and two in P2P usage are used at the aggregate level. Therefore six distributions of γ are formulated, which are reported in Figure 3 and Figure 4, respectively.

As shown in Figure 3 and Figure 4, there are considerable amounts of variance at the individual level of human dynamics both in webpage browsing and in P2P usage. To further examine if the variability on γ detected in the study does not occur by chance, variance component analysis is adopted, whose results are reported in Table 1. As it turns out, all the variance of γ are significantly different from zero at 99.9% confidence level.

The variability of γ found in the study has some implications for human dynamics research. First, as shown in Figure 3 and Figure 4, some individual γs are greater than 0, which suggest that the decision-based queuing mechanism proposed in [3] cannot be applicable to them. In other words, some individuals will not follow the decision-based queuing mechanism in task execution, while others will. Secondly, for those individuals whose γs are less than 0, the extent to which they will rely on decision-based queuing mechanism may vary across individuals. Although all of them will assign priorities to tasks before execution, they will differ in the priority index. Some individuals will assign priorities in a very extreme way – some tasks will have very high priority while the remaining will have very low priority. On the other hand, some individuals will assign priorities in a modest way – the priority they assign to tasks will gradually increase/decrease.

TABLE 1. VARIANCE COMPONENT ANALYSIS RESULTS

|  | Variance Component | Chi-square | Degrees of freedom | p-value |
|---|---|---|---|---|
| Web Browsing |  |  |  |  |
| (a) Inter-event Time | 0.0507 | 64170 | 3298 | 0.000 |
| (b) Preference | 0.0070 | 49607 | 3263 | 0.000 |
| (c) Stickiness | 0.0353 | 16513 | 3191 | 0.000 |
| (d) Diversity | 0.0064 | 3509 | 2803 | 0.000 |
| P2P Usage |  |  |  |  |
| (e) In-degree | 0.1081 | 360751 | 91088 | 0.000 |
| (f) Out-degree | 0.1498 | 192454 | 29367 | 0.000 |

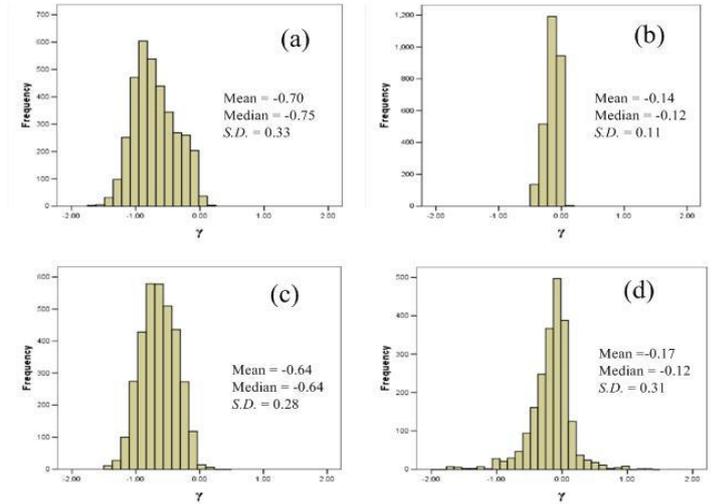

Figure 3. Distributions of γ in webpage browsing, with (a), (b), (c), and (d) corresponding to inter-event time, preference of websites, stickiness to websites, and diversity of behaviors, respectively. The mean, median and standard deviation (S.D.) of γ are reported.

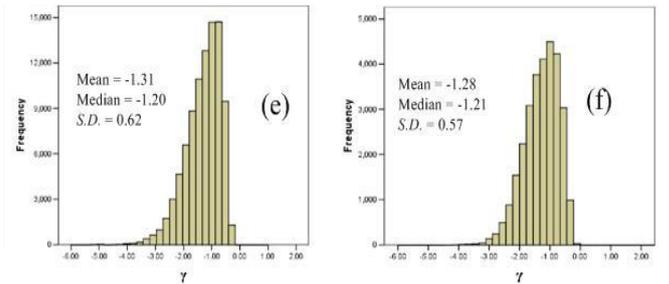

Figure 4. Distributions of γ in P2P usage, with (e) and (f) corresponding to indegree and outdegree, respectively. The mean, median and standard deviation (S.D.) of γ are reported.

*C. Distribution of γ at the Individual Level*

To investigate what kind of theoretical distribution the empirical distribution of γ in six measurements of human dynamics will follow, Kolmogorov-Smirnov test is adopted to compare the empirical distribution with three well-known theoretical distributions (i.e., Gaussian, uniform, and exponential). The results are reported in Table 2. As shown in Table 2, all the empirical distributions of γ are significantly different from these three well-know theoretical distribution except the distribution of γ in stickiness to websites measurement of webpage browsing.

To discover if other theoretical distributions can fit other five empirical distributions, probability-probability plot (P-P Plot) method is used to compare the observed γ's cumulative proportions against the cumulative proportions of four test distributions (i.e., logistic, Pareto, lognormal, and Weibull). The P-P plot results are displayed in Figure 5.

As shown in Figure 5, the distributions of γ in three measurements of human dynamics of webpage browsing (i.e., (a) inter-event time, (b) preference of websites, (c) stickiness of websites) can be well fitted by lognormal distribution,

Weibull distribution, and normal distribution (i.e., Gaussian distribution), respectively. As for the diversity of behavior of webpage browsing, the observed distribution cannot be well fitted by all test distributions in the study. As shown in Figure 5, the distribution of γ in two measurements of human dynamics of P2P usage (i.e., (e) indegree and (f) outdegree) can be well fitted by Weibull distribution.

TABLE 2 RESULTS OF KOLMOGOROV-SMIRNOV TEST

|  | Normal | Uniform | Exponential |
|---|---|---|---|
| Webpage Browsing |  |  |  |
| (a) Inter-event Time | 3.66*** | 12.29*** | 35.67*** |
| (b) Preference | 4.85*** | 12.19*** | 32.70*** |
| (c) Stickiness | 1.04 | 11.29*** | 34.30*** |
| (d) Diversity | 4.34*** | 14.15*** | 26.79*** |
| P2P Usage |  |  |  |
| (e) Indegree | 21.61*** | 156.92*** | 163.13*** |
| (f) Outdegree | 9.06*** | 63.71*** | 94.74*** |

*** The distribution differs significantly from the theoretical distribution at p < .001.

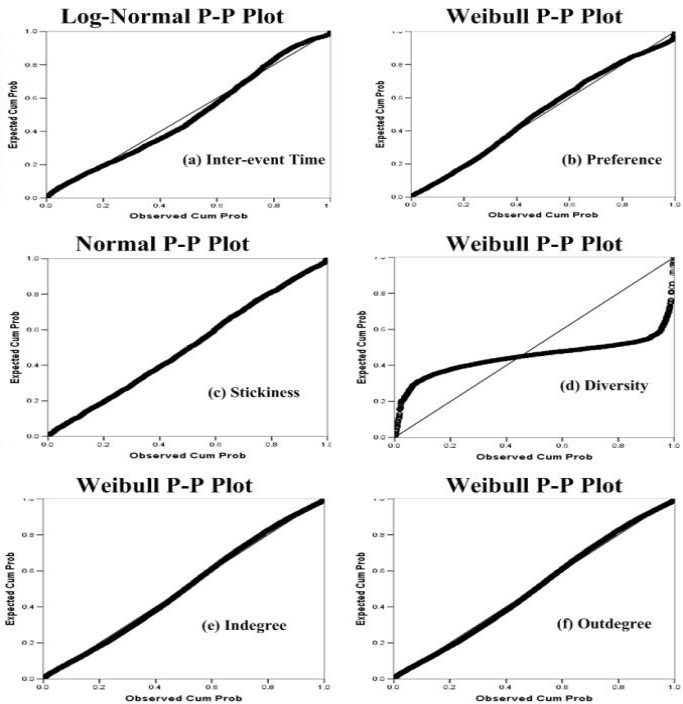

Figure 5. Probability-Probability Plot Results. The black dots are observed proportions and the straight line is the expected proportion. The observed distribution is said to match the test distribution if the black dots cluster around the straight line.

## IV. CONCLUSIONS

The study confirms the robust and ubiquitous global regularity of human communication behavior as a power-law distribution at the aggregate level. Six different operational definitions of human dynamics are examined in the study in two behavioral domains (i.e., webpage browsing and P2P usage), all of which are found to follow a power-law distribution at the aggregate level.

Although one of the behaviors we examined is webpage browsing, the exponents γ found in the study deviate from one of the two universality classes of human dynamics assuming limitations on number of tasks, which was proposed to describe web browsing in [1]. This is an empirical question that needs further exploration. One possible reason, we argue, is that in previous studies of webpage browsing [5-8] Internet users' browsing behavior within a website were investigated, while in the current study Internet users' browsing behavior across diverse websites were pooled together under investigation.

Meanwhile, the study finds a considerable amount of variability at the individual level, which does not occur by chance. This finding is very significant because it reminds us to be cautious when generalizing conclusions drawn at the aggregate level to individual phenomenon. It suggests that not all individuals follow the decision-based queuing mechanism to execute tasks. Future work is needed to explore which individual factors (e.g., demographics, life styles) can explain the variance of γ at the individual level. For those individuals who do not follow decision-based queuing mechanism, further studies are needed to explore what kind of mechanism they will adopt – random or anything else. For those individuals who follow decision-based queuing mechanism, further studies are also needed to explain their difference on γ.

Finally, it is found that the individual variability is not randomly generated, but follows several well-known families of distributions (e.g., normal, lognormal, and Weibull). Future work is needed to test the universality of distributions of γ in other behavioral domains.